\documentclass[a4paper,twocolumn]{article}
\usepackage[utf8]{inputenc}
\usepackage{hyperref}
\usepackage{graphicx}
\usepackage{color}
\usepackage{amssymb}	
\usepackage{amsmath}				
\usepackage{upgreek}	

\usepackage[paper=a4paper,left=15mm,right=15mm,top=30mm,bottom=30mm, twoside]{geometry}   

\title{\textbf{Interface effects on titanium growth on graphene}}
\author{Georg Zagler$^{1,2,*}$, Alberto Trentino$^{1,2}$, Kimmo Mustonen$^{1}$, Clemens Mangler$^1$,\\ and Jani Kotakoski$^{1,*}$\\
    $^{1}$University of Vienna, Faculty of Physics, Boltzmanngasse 5,\\ 1090 Vienna, Austria\\
    $^{2}$University of Vienna, Vienna Doctoral School in Physics, Boltzmanngasse 5,\\ 1090 Vienna, Austria\\
	$^*$Email: georg.zagler@univie.ac.at (GZ), jani.kotakoski@univie.ac.at (JK)}
\date{}
\begin{document}

\maketitle 
\begin{abstract}

    Poor quality interfaces between metal and graphene cause non-linearity and impair the carrier mobility in graphene devices. Here, we use aberration corrected scanning transmission electron microscopy to observe hexagonally close-packed Ti nano-islands grown on atomically clean graphene, and establish a 30$^{\circ}$ epitaxial relationship between the lattices. Due to the strong binding of Ti on graphene, at the limit of a monolayer, the Ti lattice constant is mediated by the graphene epitaxy, and compared to bulk Ti, is strained by ca. 3.7\%  to a value of $0.306(3)$~nm. The resulting interfacial strain is slightly greater than what has been predicted by density functional theory calculations. Our early growth stage investigations also reveal that, in contrast to widespread assumptions, Ti does not fully wet graphene but grows initially in islands with a thickness of 1-2 layers. Raman spectroscopy implies charge transfer between the Ti islands and graphene substrate.

\end{abstract}

\section*{Introduction}

The outstanding properties of two-dimensional (2D) solids such as the linear dispersion relation of charge carriers and high carrier mobility of graphene render them attractive for device applications.
However, such applications necessitate contacting the device via bulk electrodes, and results in metal-graphene interfaces that are known to influence the intrinsic properties of graphene~\cite{Lee2008a,Huard2008,Khomyakov2009}, which may reduce its usability~\cite{Venugopal2010,Moon2012}.
In addition to electronics, the interfaces of metal and graphene play a role in applications such as ultrathin catalysts~\cite{Deng2016a,Jayabal2020} and hydrogen storage~\cite{Durgun2008,Takahashi2016}. The growth of metal nanoclusters also provides a relatively straightforward way to engineer the intrinsic electronic properties of 2D materials~\cite{Jariwala2017}. Furthermore, graphene-metal composites have additionally been put forward in order to create materials with enhanced mechanical strength~\cite{Kim2013,Yang2016}.

In addition to physical and chemical vapour deposition (PVD/CVD), atomic layer deposition and wet chemistry approaches are used in the synthesis of heterostructures based on 2D materials and nanoclusters. Independent of the used growth method, however, in order to obtain surface contamination-free structures, conditions that are void of any spurious chemical species are required. This is best achieved by PVD based in ultra-high vacuum (UHV). The ratio of homogenous and defect-mediated nucleation in the PVD process are substrate dependent and hence, the choice of substrate has a substantial effect on the growth dynamics~\cite{Zhou2010}. The inert surface of 2D materials (especially that of graphene), provides a system for the study of nucleation and growth with weakly interacting substrates~\cite{Koma1992}. Indeed, as confirmed by experiments carrier out with Au and CuAu~\cite{Zagler2019,Zagler2020,Thomsen2020}, the inert surface may suppress the nucleation of nanoclusters on graphene and necessitate the use of anomalously high supersaturation to nucleate the crystal growth. These results are in a good agreement with earlier simulations that, with the exceptions of Ti~\cite{Khomyakov2009,Iqbal2012}, imply a weak interaction between graphene and metals.

Some recent reports imply that Ti may form covalent bonds with graphene and produce TiC~\cite{Gong2014}, while others have claimed that Ti cannot do so on defect-free graphene~\cite{Matruglio2016,Hsu2014}.
It has also been reported that Ti wets graphene in Frank–van der Merwe (layered) growth mode, and that it grows in a (roughly) epitaxial relationship with the graphene substrate~\cite{Hsu2014}.
Other experiments, investigating the growth of Ti on supported pristine and defective graphene, using a scanning tunnelling microscope, found that it grows in islands~\cite{Mashoff2013,Mashoff2015}.

Transport~\cite{Pi2009} and photocurrent~\cite{Lee2008a} measurements imply that significant $n$-type doping of graphene may be observable at the Ti-graphene interfaces, while Raman studies have been much less conclusive~\cite{Iqbal2012,Hsu2014}.
In one of these studies a low-yield Ti deposition onto graphene was also investigated at the atomic-scale~\cite{Zan2012a}, but due to contaminated graphene, the authors reported the Ti nucleation exclusively on the most reactive, contaminated parts of their samples. 
Nevertheless, although the structure and properties of graphene-titanium interface are still not fully resolved~\cite{Fonseca2018}, due to the strong binding of Ti on graphene and its suitable work function, Ti is being used in device experiments as an intermediate material to fabricate electrical contacts with graphene and external circuits~\cite{Moon2012}, 

Here, we report the growth of Ti on clean graphene via physical vapour deposition.
We demonstrate that Ti grows in Volmer–Weber (island) mode, resulting in clusters with hexagonal close packed (hcp) structure and a clear epitaxial relationship to the graphene substrate.
At early growth stage, Ti forms atomically thin islands that are epitaxially aligned to the graphene substrate and that have a lattice constant of $0.306(3)$~nm, corresponding to a strain of ca. 3.7\% compared to the bulk structure.
This is somewhat higher than what has been previously predicted via density functional theory calculations~\cite{Khomyakov2009,Hsu2014}.
Continuation of the growth leads to laterally larger and thicker islands, which at $5-7$ layers reach a lattice constant of $0.296(3)$~nm, close to the bulk value.
Raman spectroscopy measurements show signs of significant charge transfer.
Overall, our results bring light to the previously unexplored early stages of Ti growth on graphene, and highlight the importance of clean surfaces and atomic-scale characterization for the study of interfaces.

\section*{Experimental}
Monolayer free-standing graphene substrates were prepared from commercially available (Graphenea, ''Easy Transfer'') chemical vapour deposition-grown graphene.
Graphene, covered with a sacrificial layer (100~$\mu$m polymer film), was placed into a beaker filled with deionized water, and fished out using a perforated SiN TEM-grid (Ted Pella, 2.5~$\mu$m diameter holes).
The sacrificial layer was removed with thermal etching in an oven over 4~h at 400$^{\circ}$C in underpressure with a Ar-H gas flow.
Additional experiments were conducted using commercially available graphene-covered holey carbon membranes on gold grids (Graphenea).
The grids were baked overnight in vacuum at nominal 150$^{\circ}$C before being inserted into the interconnected ultra-high vacuum system (base pressure below 10$^{-8}$~mbar), where all subsequent experimental steps (including sample preparation and microscopy) were undertaken~\cite{Zagler2020}.
Before physical vapor deposition of Ti, the samples were further cleaned of ubiquitous contamination using a laser (ca. 450~$\mu$m spot size with powers between $10-25$~mW and exposure times between $0.5-5$~ms)~\cite{Tripathi2017}. These steps ensure that large areas free of hydrocarbon contamination and oxygen are created on free-standing graphene. On these areas, the evaporant interacts directly and exclusively with the graphene substrate, and no other experimental steps introduce oxygen that could react with the specimen.

Ti was deposited onto the specimen in ultra-high vacuum, using two different methods.
For early experiments a titanium sublimation pump was first cycled multiple times to ensure removal of possible contaminants from the evaporation material, after which Ti was deposited by placing the specimen under the pump.
For a more controlled deposition, Ti was in later experiments deposited using electron-beam physical vapour deposition from a Ti rod.
With a well-defined and better reproducible deposition rate the growth of Ti islands was observed deploying multiple deposition steps with intermediate imaging. The graphene was not heated during deposition and the deposition was performed at a base pressure of $10^{-10}$ mbar.
After deposition, the samples were transported in the interconnected vacuum system~\cite{mangler_materials_2022} to the Nion UltraSTEM 100 scanning transmission electron microscopy (STEM) instrument operated here at 60~kV.
The convergence semiangle was 30~mrad, and images were acquired with medium- and high-angle annular dark-field detectors (MAADF and HAADF, annular ranges of 60--200~mrad and 80--280~mrad, respectively).
Electron energy-loss spectroscopy (EELS) was carried out in the same microscope using a Gatan PEELS 666 spectrometer fitted with an Andor iXon 897 electron-multiplying charge-coupled device camera~\cite{Susi2017}.
STEM image simulations were performed using AbTEM ~\cite{Madsen2021}.

The samples, pre-characterized with Raman spectroscopy, were cleaned with the laser as described above, with each free-standing graphene patch (covering a hole of the TEM-grid) having between 10 and 60\% of the surface free of contamination, as established via STEM imaging.
After pre-characterization, Ti was deposited using the sublimation pump.
In contrast to the STEM measurements, Raman spectroscopy was conducted outside the vacuum system, transferring the specimen in air, but minimizing the air exposure to a maximum of 3~h.
For collecting the spectra, a 532~nm laser with 5~mW of power, 0.5~s exposure time, 30 accumulations and a 600~gr/mm grating were used to measure the Raman response. D, G and 2D peaks were fitted with Lorentzian peaks.

\section*{Results and Discussion}

\begin{figure*}[h]
	\includegraphics[width=\textwidth]{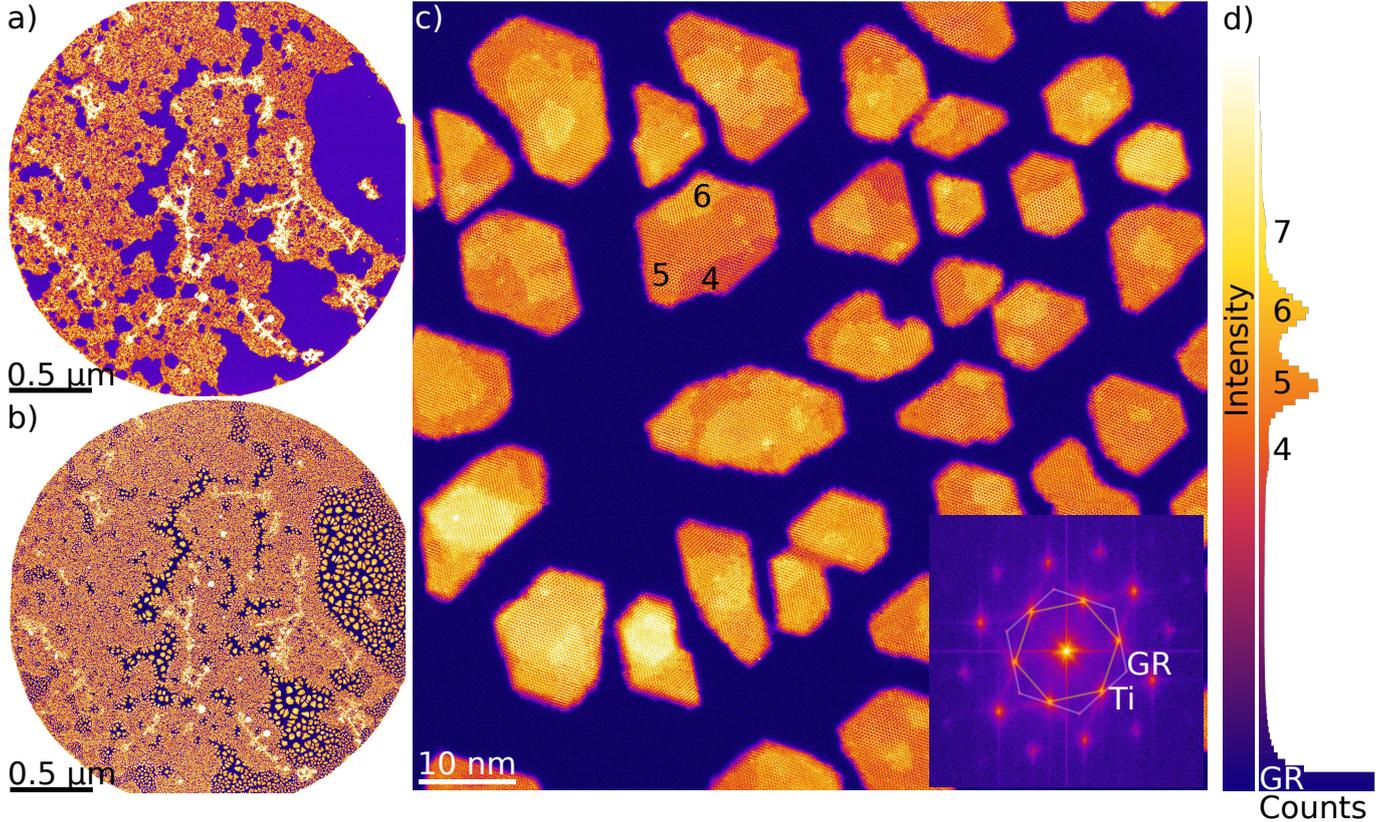}
	\caption{\textbf{Ti clusters on graphene and contamination.}
    \textbf{a)} STEM HAADF pre-characterization image of suspended graphene after laser cleaning.
    In brighter areas surface contaminants cover the otherwise clean graphene.
    \textbf{b)} STEM HAADF image of titanium deposited onto graphene, where it grows large, mostly uniform islands on the clean sample areas.
    \textbf{c)} STEM HAADF image of Ti islands showing layered growth.
    Fourier transform (inset) reveals the hexagonal crystal structure of the Ti islands and the epitaxial relationship with the graphene substrate at 30$^{\circ}$ misalignment.
    The corresponding intensity histogram (after a Gaussian blur of the image in panel \textbf{d}) shows peaks corresponding to a thickness of $4-7$ layers.
    }
	\label{fig:Overview} 
\end{figure*}

STEM MAADF images in Fig.~\ref{fig:Overview} show the growth of Ti on free-standing monolayer graphene.
Before deposition (Fig.~\ref{fig:Overview}a), the specimen is cleaned with the laser, which creates large areas (blue, some 100~nm in diameter) of atomically clean graphene.
After Ti deposition (Fig.~\ref{fig:Overview}b), changes in both the contamination-free and contaminated graphene can be seen.
On contamination (not shown), Ti grows in small connected clusters that display no uniform orientation or lattice parameter, most likely due to the strong interaction between Ti and C in the contamination that unlike C in graphene do not have ideal bonding configuration.
In contrast, previously clean graphene areas are now covered with Ti structures with a uniform size distribution, whereas contaminated areas show a more uniform increase in contrast.
In Fig.~\ref{fig:Overview}c a STEM HAADF image encloses several clusters on clean graphene and reveals the actual morphology of the deposited Ti structures.
The clusters show a hexagonal crystal structure, as is apparent in the Fourier transform shown in the inset (bottom right corner). The color of the clusters changes with discrete steps, showing terraces.
In the STEM HAADF images, the contrast (that defines the color depicted here) can be directly interpreted as a thickness.
The histogram of the pixel intensities (after a Gaussian blur with a $\sigma$ of seven pixels) of Fig.~\ref{fig:Overview}c, depicted in Fig.~\ref{fig:Overview}d, confirms discrete steps in the intensities. 
This means that the Ti islands display terraces with clearly defined atomic layers (with a typical thickness of ca. 5--6 layers in the case of this image).
As evident from the Fourier transform, the Ti island are epitaxial with the graphene substrate with a 30$^\circ$ misorientation, and have a lattice constant of 0.296(3)~nm, close to the bulk value of 0.295~nm~\cite{Wood1962}.

\begin{figure*}[h]
	\includegraphics[width=\textwidth]{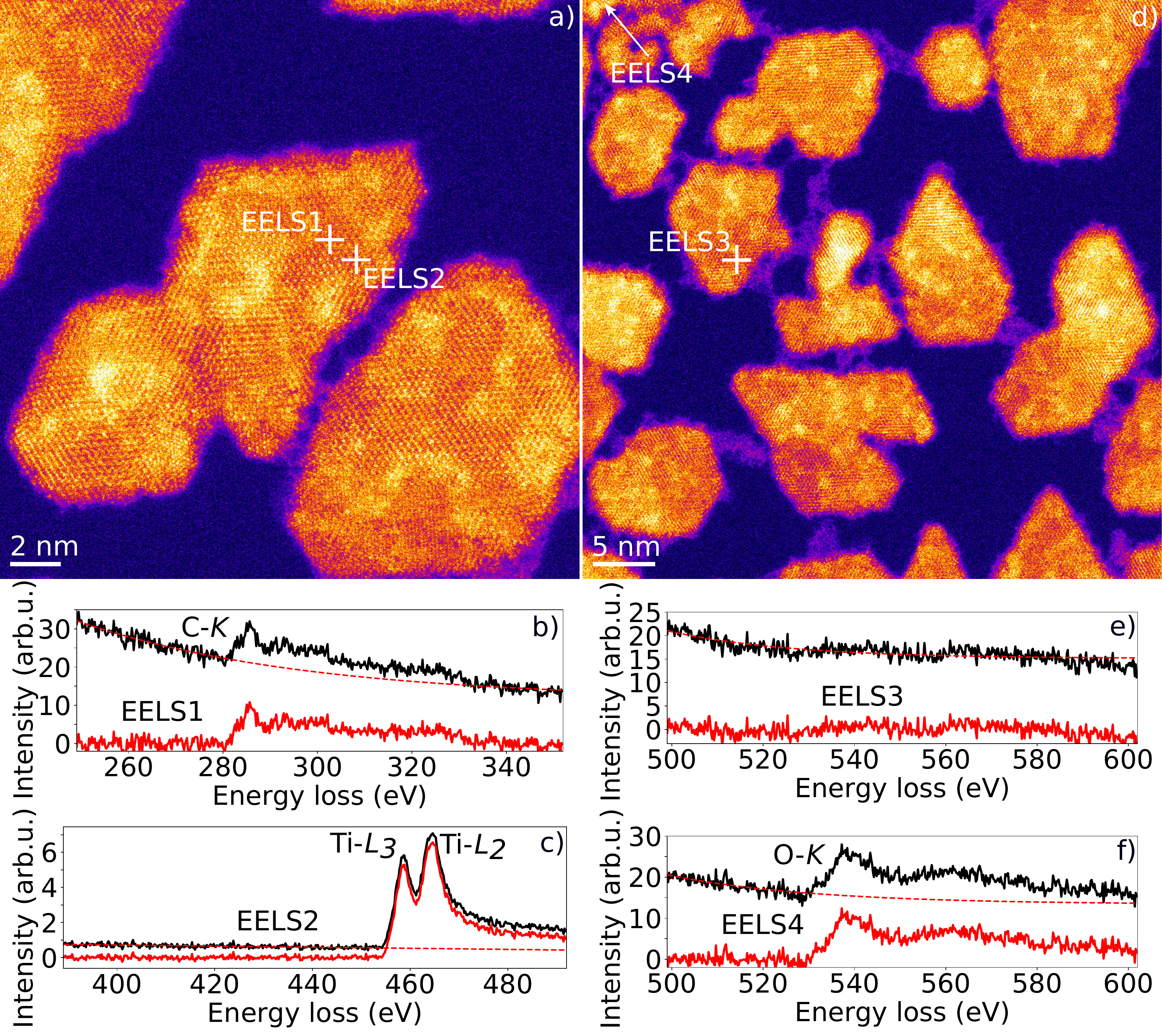}
	\caption{\textbf{Electron energy-loss spectroscopy (EELS) of Ti clusters on graphene.}
    \textbf{a)} STEM HAADF image of Ti islands on graphene showing the locations where EELS spectra (EELS1 and EELS2) were recorded.
    EEL spectra EELS1 in panel \textbf{b} and EELS2 in panel \textbf{c} show carbon K-edge and Ti L-edges from the Ti recorded on a Ti island, respectively.
    The background subtracted signal (red solid line) is obtained by subtracting an exponential fit function (red dashed line) from the experimental signal (black solid line).
    \textbf{d)} STEM HAADF image of Ti islands on graphene showing the locations where EELS spectrum EELS3 (shown in panel \textbf{e}) was recorded.
    The arrow points to the location (slightly outside of the figure) where spectrum EELS4 (shown in panel \textbf{f}) was recorded on area containing some contamination that was left over after the sample was cleaned.
    The background subtracted signal (red solid line) is obtained by subtracting an exponential fit function (red dashed line) from the experimental signal (black solid line).
    }
	\label{fig:EELS} 
\end{figure*}

Electron energy loss spectroscopy (EELS) was carried out on the structures after the deposition.
EEL spectra of a Ti island (Fig.~\ref{fig:EELS}a,c) confirm the presence of Ti by the characteristic Ti $L_2$ and $L_3$ core-loss edges.
Signal from the carbon $K$-edge, stemming from the graphene support, can also be observed at the position of the Ti islands (Fig.~\ref{fig:EELS}a,b).
The carbon $K$-edge can be used for chemical analysis of the bonding, since while $sp^2$-hybridized carbon has two distinct peaks, $\pi ^*$ and $\sigma ^*$, $sp^3$-hybridized carbon only shows a single $\sigma ^*$ peak~\cite{Hamon2004}.
Here the weak carbon $K$-edge signal does not show a distinct $\pi ^*$ peak followed by $\sigma ^*$, suggesting that the Ti interacts strongly with the graphene support, promoting  $sp^3$-hybridization.
In Figure~\ref{fig:EELS}d,e,f the energy-loss region containing the oxygen $K$-edge is shown from two different positions.
On the Ti island (EELS3), no sign of the oxygen $K$-edge can be found, confirming the lack of oxygen, as can be expected for evaporation carried out in ultra-high vacuum.
In contrast, the spectrum recorded on the contamination remaining after laser cleaning, next to the atomically clean area with Ti islands, the characteristic oxygen $K$-edge can be found (Fig.~\ref{fig:EELS}f).

We stress that the Ti islands shown here have the same hexagonal close packed structure as bulk Ti (where one layer consists of two atomic layers with AB stacking), and that structures grown on the carbon-based contamination and under oxygen-containing conditions lead to a different atomic structure.
Moreover, the atomically clean graphene around the Ti islands remains pristine.
Therefore, it is clear that these Ti clusters consist indeed of pure titanium grown on atomically clean graphene.

\begin{figure*}
	\includegraphics[width=\textwidth]{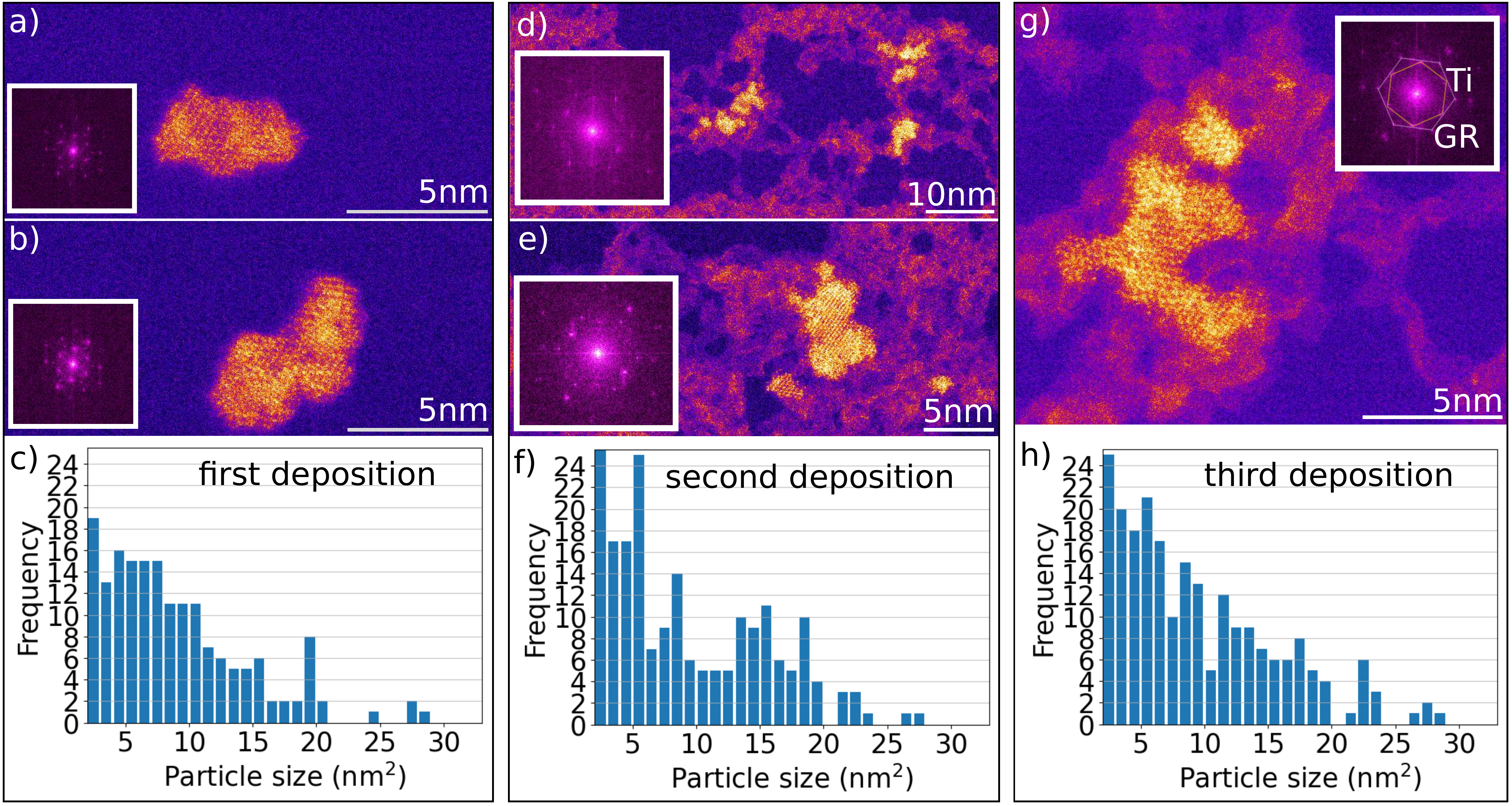}
	\caption{\textbf{Ti nanocluster growth with step-wise deposition.}
    \textbf{a-b)} STEM HAADF example images of Ti islands after the first deposition step.
    They have a hexagonal crystal structure with some monolayer areas (single-atom contrast) and some thicker areas.
    Insets are the Fourier transforms of each image, confirming the hexagonal structure (hexagonal pattern with broadened peaks) and the graphene substrate (sharp hexagonal pattern).
    \textbf{c)} Island size distribution after the first deposition step.
    \textbf{d-e)} STEM HAADF example images after the second deposition step.
    Some clusters still have some monolayer areas, while many are covered with another layer, which typically does not display the perfect hexagonal structure.
    \textbf{f)} Size distribution after the second deposition step.
    \textbf{g)} STEM HAADF example image after the third deposition step.
    \textbf{h)} Size distribution after the third deposition step.
    }
	\label{fig:Growth_early}
\end{figure*}

To better understand the growth, low quantities of Ti were next deposited in three steps.
At low amounts, small Ti clusters grow on clean defect-free graphene (Fig.~\ref{fig:Growth_early}).
The clusters have an average size of 9.2 nm$^2$ (logarithmic distribution) and display monolayer (ML) areas with a hexagonal structure (see Fourier transform in the inset, sharp peaks correspond to graphene, broader peaks to Ti), as well as some Ti atoms on the ML region.
These atoms can form a second or a third layer (Fig.~\ref{fig:Growth_early}b), but are not always perfectly aligned (Fig.~\ref{fig:Growth_early}a).
After the first deposition, the coverage is 0.7\% with 700~islands/$\mu$m$^2$ of clean graphene.
It should be noted that all clusters also align with the graphene lattice, and no exclusively monolayer Ti island was found.

The same specimen is then deposited with more Ti (Fig.~\ref{fig:Growth_early} d-f). Contamination visible in the images (red in color, covering large parts of the graphene) is due to mobile contamination being pinned down by the electron beam during imaging~\cite{Zagler2019}.
The island growth is not influenced by this, as the contamination only appears after deposition during imaging, and after each deposition step, different sample area was imaged.
The average cluster size is 9.4~nm$^2$, and many clusters appear to have more layers.
At the same time the coverage increases to 1.8\% and the cluster density increases to 1900~islands/$\mu$m$^2$.
The distribution of the clusters is similar to the first deposition (Fig.~\ref{fig:Growth_early}c), with a small increase in the number of both larger and smaller clusters.
The Fourier transform still displays the same hexagonal pattern with epitaxial alignment between the islands and the graphene substrate.

A consecutive Ti deposition on the same specimen (Fig.~\ref{fig:Growth_early} g-h) yields similarly sized Ti clusters (8.4 nm$^2$ on average) with a distribution (Fig.~\ref{fig:Growth_early}h) similar to the first deposition (Fig.~\ref{fig:Growth_early}c).
The coverage (1.5\%) as well as the density (1800~islands/$\mu$m$^2$) are comparable to the second deposition.
It should be noted that some pinned-down hydrocarbon contamination was already present and appears to have not altered the growth significantly (if it did, one would expect to find many smaller disordered clusters similar to what was described above for contaminated sample areas).
This is in line with previous experiments that have shown that thin contamination coverage does not necessarily act as a preferred nucleation site~\cite{Zagler2019}.
The growth in 3D islands at lower and step-wise deposition confirms the Volmer-Weber island-type growth-mode.


A feature common to many multilayer Ti nanoislands is an apparent lattice mismatch within the hexagonal structure, seen in all Ti island (see for example the island marked with numbers in Fig.~\ref{fig:Overview}c).
We assume that this is due to the strong interaction between graphene and the first Ti layer which imposes strain on the structure, but decreases layer by layer as the distance from graphene increases.
This is in line with simulations that predict strain at the Ti-graphene interface~\cite{Hsu2014}.
Specifically, Hsu and co-workers~\cite{Hsu2014} used experimental input to create a Ti-graphene superstructure and calculated a 2\% interface strain that was assumed to be in graphene.
In the same study, the authors also reported a substantial shift in the Raman G-peak of graphene, which can be attributed to strain in graphene caused by a Ti layer.

\begin{figure*}
	\includegraphics[width=\textwidth]{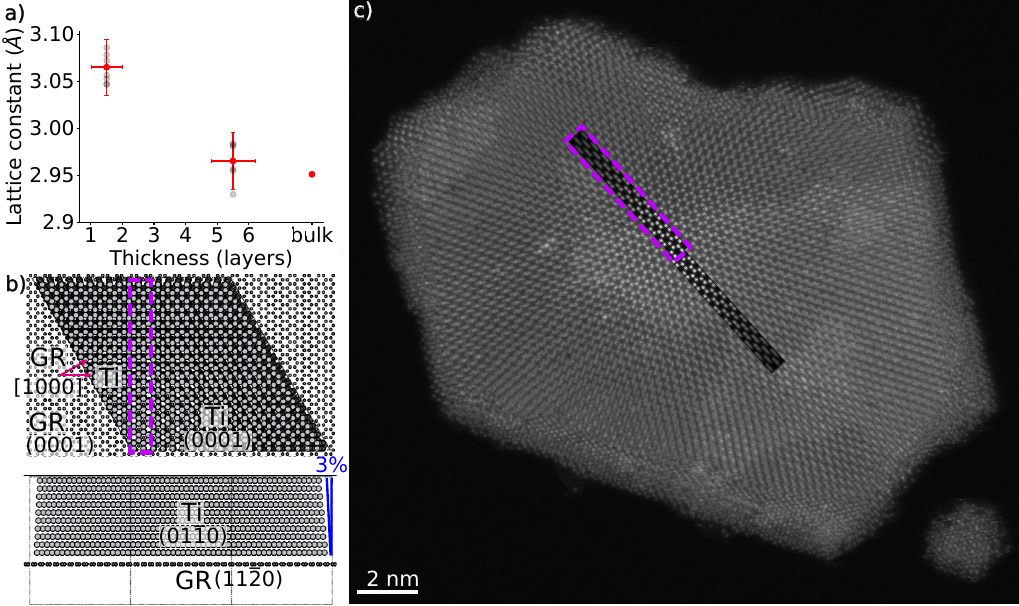}
	\caption{\textbf{Strain in Ti nanoclusters on graphene substrate.}
    \textbf{a)} Lattice constant of Ti islands decreases by 3.4\% from 0.306(3)~nm ($1-2$ layers) to 0.296(3)~nm ($5-6$ layers) with increasing island thickness.
    The red points are means calculated from all experimental images (gray points), and the error bars correspond to the standard deviation of the data and maximum uncertainty in determining the number of layers from the STEM HAADF image contrast.
    \textbf{b)} Atomic model of a Ti slab consisting of five Ti layers (each consisting of two atomic layers in AB-stacking) with 3\% linear change in the lateral lattice constant (i.e. strain in (0001) plane) from the lowest to the highest layer. The free-standing monolayer graphene substrate is depicted with a 30$^{\circ}$ misalignment to the Ti slab.
    \textbf{c)} STEM HAADF image of Ti island on graphene overlaid with a simulated image from the 5 layer Ti slab with strain shown in panel b from the area marked with a rectangle.
    }
	\label{fig:Strain}
\end{figure*}

To investigate this further, we evaluated the Ti lattice constant from the Fourier transform, by comparing it to the lattice constant of graphene estimated from the corresponding sharp peaks.
Interestingly, contrasting the lattice constant of thin regions (Fig.~\ref{fig:Growth_early}a) to mostly $5-6$-layered islands (Fig.~\ref{fig:Overview}c), reveals that there is a significant strain in the thin Ti islands.
Indeed, the lattice constant in the thinnest structures is about 0.306(3)~nm, whereas in the thicker islands it is 0.296(3)~nm, that is close to bulk Ti in $\alpha$-phase at 0.295~nm~\cite{Wood1962}.
This corresponds to a strain of ca. 3.7\%.
The data is shown in Fig.~\ref{fig:Strain}a.
To confirm our hypothesis that the appearance of the thicker structures could be explained by gradually changing strain within the Ti islands, we created an artificial Ti structure with five AB layers (ten atomic layers), where the in-plane lattice constant is varied by 3\% linearly between the first and the last layer. The corresponding structure of Ti islands, including the monolayer graphene substrate, with variation in the Ti in-plane lattice constant, is depicted in Fig.~\ref{fig:Strain}b. A variation in the out-of-plane distance between hexagonal Ti layers is omitted in the model, as it would neither influence the image simulation, nor could it be observed in the STEM data.
In Fig.~\ref{fig:Strain}c an image simulation of the part of the Ti structure highlighted with a rectangle in Fig.~\ref{fig:Strain}b is overlaid on top of an experimental STEM HAADF image of an island with a comparable thickness.
As can be immediately recognized, the simulated image is in excellent agreement with the experimental one confirming our hypothesis.

\begin{figure*}
	\includegraphics[width=\textwidth]{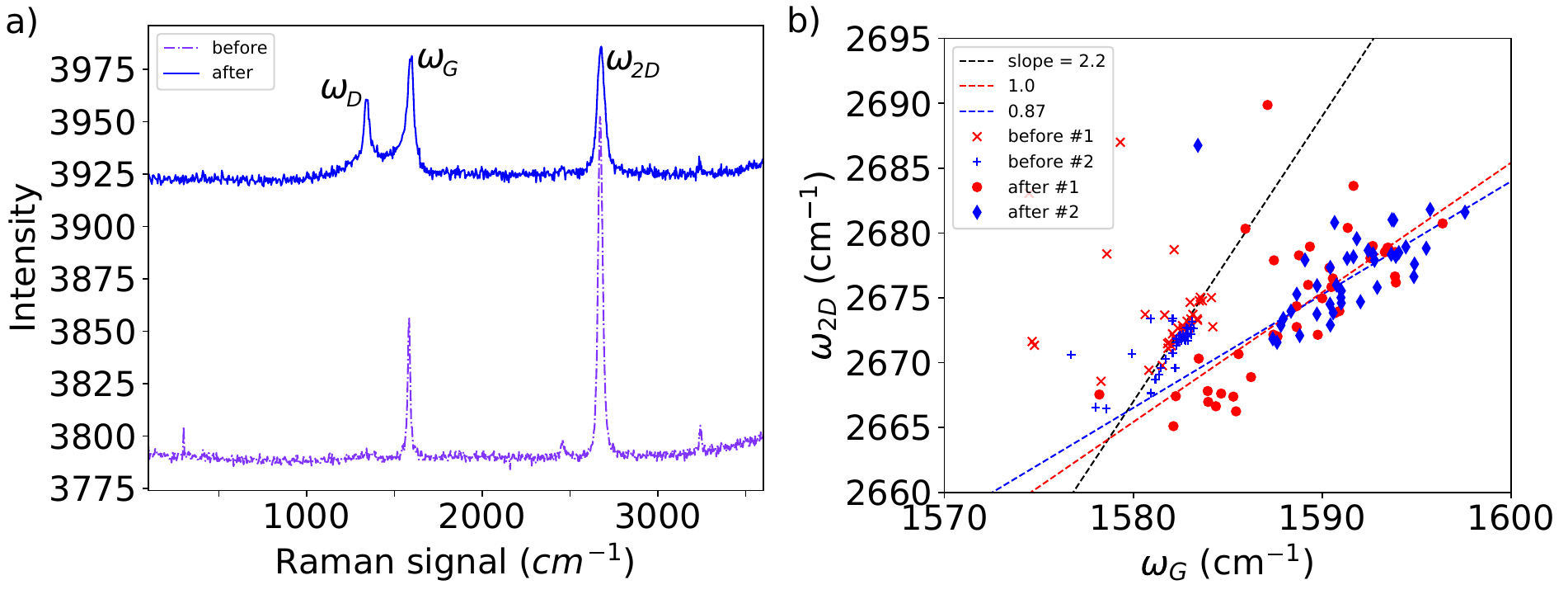}
	\caption{\textbf{Raman spectra of graphene before and after Ti deposition.}
    \textbf{a)} Example Raman spectra before laser cleaning and Ti deposition, and after (shifted in intensity).
    \textbf{b)} Raman G and 2D peak positions for two samples (\# 1 and \# 2). Before Ti deposition both samples show a linear dependency between the peaks with an approximate slope of 2.2 (black dashed line).
    After Ti deposition the peaks show another linear distribution with slopes of 1.0 and 0.87, respectively.
    }
	\label{fig:Raman}
\end{figure*}

Raman spectroscopy was conducted before loading the specimen into the vacuum system, after laser cleaning (10-60\% of the graphene surface cleaned), and after Ti deposition.
The main goal of these experiments was to assess the quality of graphene before the evaporation, and to provide information on possible charge transfer between graphene and the Ti islands.
Most suspended areas covered with Ti islands were 20-40\% free of surface contamination (with direct contact between Ti and graphene).
The pristine free-standing graphene has a Raman response with a G peak at 1584.1~cm$^{-1}$ and a 2D peak at 2670.1~cm$^{-1}$ (Fig.~\ref{fig:Raman}a).
The peaks are sharp and no D peak is present.
After laser cleaning and Ti deposition both G and 2D peaks are shifted up (to 1592.0~cm$^{-1}$ and 2675.9~cm$^{-1}$, respectively).
The peaks are also broadened, and the D peak appears.
The $I_D/I_G$ signal intensity ratio after deposition is ca. 0.31~$\pm$ 0.21, which indicates a change in the graphene substrate, probably due to the strong interaction with the Ti islands.

The G and 2D peaks show a linear dependency (Fig.~\ref{fig:Raman}b) for the as-prepared samples with an approximate slope of 2.2.
While the linearity remains after laser cleaning and Ti deposition, the slope decreases significantly down to values close to 1 and below (depending on the sample).
Slopes close to 2.2 are well-established to arise from uniaxial strain in graphene~\cite{Mohiuddin2009,Elibol2016, Hsu2014,Lee2012a}, whereas values close to 0.75 have been attributed to $p$-type doping~\cite{Lee2012a}.
For $n$-type doping, a non-linear global behavior is expected, although it would appear nearly linear for the range of values investigated here~\cite{Das2009,Lee2012a}.
We point out that for our samples, the situation is complicated by the laser cleaning step, which can lead to either negative or positive strain in the sample due to thinning and possibly partial cracking of the sample support.
Other influence arising from heating, such as hole doping or substrate interactions~\cite{Yoon2011, Lee2012a} should however be absent for suspended monolayer graphene samples with a holey SiN support, laser heated in high vacuum.

Finally, it is necessary to point out that the Ti islands only cover a fraction of the area under the laser spot used for Raman spectroscopy.
Therefore, the results can not be exclusively interpreted to correspond to the interface between graphene and the Ti island, but instead provide an average signal from different types of areas of the sample (including atomically clean graphene with Ti islands, but also areas with contamination and possibly structures combining Ti and C).
Nevertheless, the observed change in the G/2D peak behavior is a clear indication of charge transfer in our Ti deposited samples.
Significant widening of the G peak or sharp decreases in the 2D peak intensities that have been previously reported~\cite{Hsu2014,Iqbal2012} were not observed in our samples.

\section*{Conclusion}

Ti deposited onto demonstrably clean, monolayer graphene nucleates on clean sample areas already at low densities and grows in the Volmer–Weber (island) fashion.
The resulting Ti islands display the expected hcp crystal structure with different layers and step edges clearly visible.
For the thinnest structures directly in contact with graphene, the formation of an epitaxial interface with a 30$^\circ$ misorientation leads to a significant strain of up to over 3\%, and a corresponding lattice constant of 0.306(3)~nm.
By island thickness of ca. 5--6 AB layers (10--12 atomic layers) or more, the lattice constant decreases to 0.296(3)~nm ($<0.5$\% strain compared to bulk value).
Based on Raman spectroscopy, Ti deposited onto graphene leads to clear charge transfer independent of surface contamination.
In contrast to island growth on clean graphene, Ti deposited onto contaminated graphene grows as non-crystalline clusters.
These results highlight how important characterization of graphene and removal of the ubiquitous surface contamination is for the complete understanding of the interface effects between graphene and other materials.

\section*{Acknowledgments}

We acknowledge the Austrian Science Fund (FWF) for funding through project P31605.

\bibliography{Bibliography2}

\noindent {\bf Competing interests:} The authors declare no competing interests.

\noindent {\bf Data availability:} All data needed to evaluate the conclusions in the paper are present in the paper or the supplementary materials.


%
%

\end{document}